\documentclass[12pt]{amsart}
\usepackage[square,compress,comma, numbers,sort]{natbib}
\usepackage[colorlinks=true, citecolor=blue, linkcolor=blue]{hyperref}
\usepackage{amsfonts}
\allowdisplaybreaks[4]
\usepackage{graphicx}
\usepackage{epstopdf, booktabs}
\usepackage{subfigure}
\usepackage{caption}
\usepackage[svgnames]{xcolor}
\usepackage{amssymb}
\usepackage{amsmath, url}
\usepackage{color}
\usepackage{enumerate}
\usepackage{float}
\usepackage{mathtools, multirow}

\usepackage{listings}
\lstdefinestyle{mystyle}{
  backgroundcolor=\color{backcolour}, commentstyle=\color{codegreen},
  keywordstyle=\color{magenta},
  numberstyle=\tiny\color{codegray},
  stringstyle=\color{codepurple},
  basicstyle=\ttfamily\footnotesize,
  breakatwhitespace=false,         
  breaklines=true,                 
  captionpos=b,                    
  keepspaces=true,                 
  numbers=left,                    
  numbersep=5pt,                  
  showspaces=false,                
  showstringspaces=false,
  showtabs=false,                  
  tabsize=2
}

\usepackage{threeparttable}

\newcommand{\pk}[1]{\mathbb{P} \left\{ #1 \right\} }

\newcommand{\expon}[1]{\exp\left(#1\right)}

\newcommand{\R}{\mathbb{R}}
\newcommand{\EE}{\mathbb{E}}
\newcommand{\inr}{\in \R}

\newcommand{\COM}[1]{}

\newcommand{\kb}[1]{\boldsymbol{#1}}
\newcommand{\vk}[1]{\kb{#1}}

\def\E#1{\mathbb{E}\left \{#1 \right\}}

\def\d{\,\mathrm{d}} % For the integral
\def\I#1{\mathbb{I}\left (#1 \right)}

\def\N{\mathbb{N}}
\def\IF{\infty}

\def\VaR{\mathrm{VaR}}
\def\CVaR{\mathrm{CVaR}}

\definecolor{codegreen}{rgb}{0,0.6,0}
\definecolor{codegray}{rgb}{0.5,0.5,0.5}
\definecolor{codepurple}{rgb}{0.58,0,0.82}
\definecolor{backcolour}{rgb}{0.95,0.95,0.92}

\definecolor{c20}{rgb}{0.,0.7,0.}
\definecolor{c30}{rgb}{0.,0.,1.}
\definecolor{c40}{rgb}{1,0.1,0.7}
\definecolor{c50}{rgb}{1,0,0}
\definecolor{c60}{rgb}{1,0.9,0.1}
\definecolor{c70}{rgb}{0.50,1.00,0.00}

\def\cL#1{\textcolor{c50}{#1}}
\def\cL#1{#1}
\def\CCL#1{\textcolor{c30}{#1}}
\def\CCL#1{#1}
\def\CL#1{\textcolor{c30}{#1}}
\def\CL#1{#1} %uncomment to remove the color label, you may define your personal color to highlight those changes
\def\LE#1{\textcolor{c40}{#1}}
\def\LP#1{\textcolor{c50}{#1}}

\def\LE#1{#1}\def\LP#1{#1}

\numberwithin{equation}{section}
\newtheorem{theo}{Theorem}[section]
\newtheorem{sat}[theo]{Proposition}
\newtheorem{de}[theo]{Definition}
\newtheorem{lem}{Lemma}[section]

\newtheorem{korr}[theo]{Corollary}

\newtheorem{remarks}[theo]{Remarks}
\numberwithin{equation}{section}

\newcommand{\BQN}{\begin{eqnarray}}
\newcommand{\EQN}{\end{eqnarray}}
\newcommand{\BQNY}{\begin{eqnarray*}}
\newcommand{\EQNY}{\end{eqnarray*}}

\newcommand{\BS}{\begin{sat}}
\newcommand{\ES}{\end{sat}}
\newcommand{\BT}{\begin{theo}}
\newcommand{\ET}{\end{theo}}
\newcommand{\BK}{\begin{korr}}
\newcommand{\EK}{\end{korr}}

\newcommand{\BD}{\begin{de}}
\newcommand{\ED}{\end{de}}
\newcommand{\BIT}{\begin{itemize}}
\newcommand{\EIT}{\end{itemize}}
\newcommand{\BDI}{\begin{description}}
\newcommand{\EDI}{\end{description}}
\newcommand{\BRM}{\begin{remarks}}
\newcommand{\ERM}{\end{remarks}}
\newcommand{\BEL}{\begin{lem}}
\newcommand{\EEL}{\end{lem}}
\date{}

\begin{document}
	
\title{Extremal Analysis of Flooding Risk and Management}%{Extreme Analysis of Flood Disaster Risks in China with Insurance and Financial Management}
%\footnote{Chengxiu Ling would like to thank Prof. D\c{e}bicki Krzystof   for his nice suggestions during  visit of University of Wroc\l aw, Poland.}	

\author{Chengxiu Ling}
	\address{Academy of Pharmacy, Xi'an Jiaotong-Liverpool University, Suzhou, SIP 215123}
	\email{Chengxiu.Ling@xjtlu.edu.cn}

\author{Jiayi Li}
     \address{Department of Financial and Actuarial Mathematics, Xi'an Jiaotong-Liverpool University, Suzhou, SIP 215123}
     \email{Jiayi.Li1802@student.xjtlu.edu.cn}
	
\author{Yixuan Liu}
     \address{Department of Financial and Actuarial Mathematics, Xi'an Jiaotong-Liverpool University, Suzhou, SIP 215123}
     \email{Yixuan.Liu18@student.xjtlu.edu.cn}
	
\author{Zhiyan Cai}
     \address{Department of Biology Information, Xi'an Jiaotong-Liverpool University, Suzhou, SIP 215123}
     \email{Zhiyan.Cai18@student.xjtlu.edu.cn}

{
	\bigskip
	
	\date{\today}
	\maketitle

\begin{abstract}
	\footnotesize{Catastrophic losses caused by natural disasters receive a growing concern about the severe rise in magnitude and frequency. The constructions of insurance and financial management scheme become increasingly necessary to diversify the disaster risks. Given the frequency and severity of floods in China, this paper investigates the extreme analysis of flood-related huge losses and extreme precipitations using Peaks-Over-Threshold method and Point Process (PP) model. These findings are further utilized for both designs of flood \CCL{zoning} insurance and flooding catastrophic bond: (1) Using the extrapolation approach in Extreme Value Theory (EVT), the estimated Value-at-Risk (VaR) and conditional VaR (CVaR) are given to determine the cross-regional insurance premium together with the Grey Relational Analysis (GRA) and the \CCL{Technique for Order Preference by Similarity to an Ideal Solution (TOPSIS)}.  The flood risk vulnerability and threat are analyzed with both the geography and economic factors into considerations, leading to the three layered premium \CCL{levels} of the 19 flood-prone provinces. (2) To hedge the risk for insurers and reinsurers to the financial market, we design a flooding catastrophe bond with considerate trigger choices and the pricing mechanism to balance the benefits of both reinsurers and investors. To reflect both the market price of catastrophe risk and the low-correlated financial interest risk, we utilize the pricing mechanism of Tang and Yuan (2021) to analyze the pricing sensitivity against the tail risk of the floods’ disaster and the distortion magnitude 
and the market risk through the distortion magnitude involved in Wang's transform. Additionally, our trigger process is carefully designed using a compound Poisson process modelling both the frequency and the layered intensity of the flood disasters.  Finally, constructive suggestions and policies are proposed concerning the flood risk warning and prevention.
	
 }

\end{abstract}

\newpage

\section{Introduction}\label{introduction}
Extreme weather events \cL{pose} great threats to human lives and cause significant financial losses. Climate changes in recent decades \cL{increase} the occurrence and the severity of such events (\citep{Chen2013}). Extreme precipitations, as one of the extreme weather events, usually come along with management threats such as severe floods. \cL{In} Asian monsoon season, the precipitations usually will trigger floods, especially in the basin of the largest river in China (\citep{Kundzewicz2020}, \citep{tang_yuan_2019}). The historic large floods in China in the past three decades have caused over 200 billion dollars per decade (\citep{Kundzewicz2020}). 
In August, 2021, the devastating floods in the Chinese city of Zhengzhou saw 457.5 millimeters of precipitation within 24 hours, which caused severe flooding and a resulting-in RMB 53.2 billion economic losses and more than RMB 6.4 billion in insurance claims resulted from over \cL{$4\times 10^5$ cars} damages. Such extreme severe disaster occurs so frequently all over the world.  \cL{Despite hazard mitigation efforts and scientific and technological advances, extreme weather events continue to cause substantial losses.  The concern of this paper is to develop quantitatively an insurance and financial risk management scheme to diversify the high layers of flood risks. }

Extreme Value Theory (EVT) is a useful tool to evaluate and simulate extreme data of such kind of weather events, \cL{as well as financial crises, super-spreading of contagious diseases e.g., COVID-19, severe rainstorms etc (\citep{cirillo2020tail}, \citep{embrechts2013modelling},  \citep{falk2011extreme}, \citep{towler2020extreme}). The mis-specifications of tail risks might result in underestimating potential risks, causing the mighty breakdowns of health care systems involved. }
There are generally two methods to analyze extremes. One is the Block Maxima (BM) approach, which extracts maximum values of each block from the time series.
\cL{Another approach} 
is the \cL{P}eaks-\cL{O}ver-\cL{T}hreshold (POT) approach, which analyzes threshold-excess by generalized Pareto (GP) models. By Fisher-Tippett theorem, the Block Maxima data is usually fitted with Generalized Extreme Value (GEV) distribution. Both methods have a common challenge to overcome the distribution uncertainty when identifying extremes as well as the rareness of extreme data, see \citep{falk2011extreme, Gilleland2016}.

In this paper, we focus on the risk management of flooding risk in China based on the economic loss and precipitations in 2006-2018 following the workflow in Figure \ref{workflow} below. 
First, it turns out that both possess extreme features fitted well by extreme value models, see Section \ref{sec: EVA}. Given the various climate and topography imposed to the flood-prone regions in China together with different economic development, we implement Multiple-Criteria Decision-Making methods (\citep{Deng1982, Zhao2013, Ho2006, Sun2020}) to design the layered and cross-regional insurance compensation scheme to diversify the extreme flooding risk in the framework of extreme statistics and risk management in Section \ref{cross-regional}. Additionally, concerned with the financial loss caused by super severe flood disasters, we consider the risk mitigation from the reinsurance to  financial markets by designing catastrophe (CAT) bond as a financial instrument of re-insurance in Section \ref{bond}, so that financial resilience can be increased (\citep{Chen2013, Zhou2017, Surminski2014}). Indeed, the Chinese financial market is ready to issue the CAT bond in the Hong Kong market through special-purpose insurance companies to transfer catastrophe risk loss, 
according to the China Banking and Insurance Regulatory Commission.  The issuer can exempt or delay the payment of part of the bond principal and interest or even all of them when actual catastrophe loss exceeds the agreed amount. 
\begin{figure}[H]
    \centering
    \includegraphics[scale = 0.5]{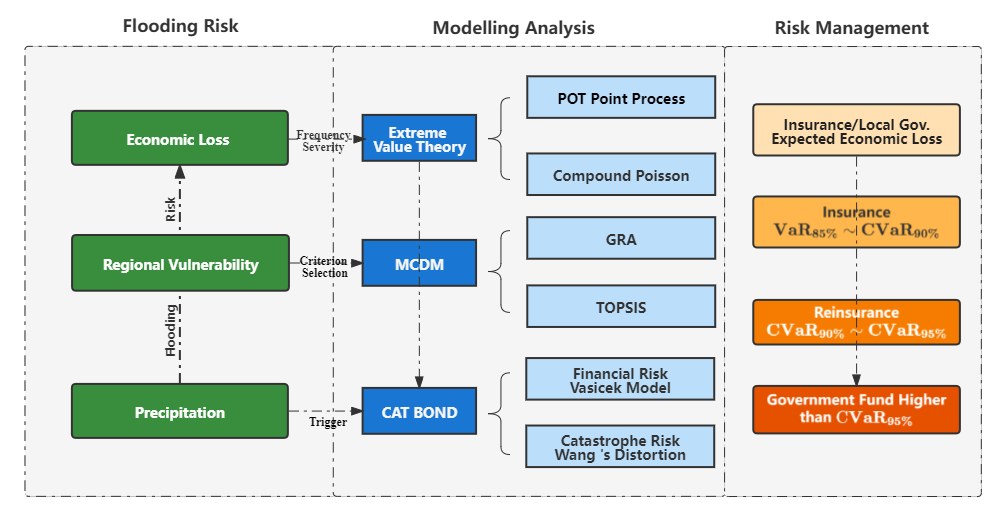}
    \caption{Workflow of flooding risk management.}
    \label{workflow}
\end{figure}

The main contributions of the paper are three-fold. 
\begin{itemize}
\item The layered compensation insurance scheme is well designed according to the estimated Value-at-Risk (VaR) and conditional VaR (CVaR), incorporating the 
extrapolation method in extreme value theory, see Table \ref{tab: VaR-CVaR}. This provides policy-makers constructive suggestions for allocating abundant monetary support and flood risk prevention among all stakeholders.   
\item In the design of cross-regional insurance fund, the sensitivity of criterion selection and weight is studied in terms of  Multiple-Criteria Decision-Marking (MCDM) method \CL{using} the Grey Relational Analysis (GRA) and the Technique for Order Preference by Similarity to an Ideal Solution (TOPSIS). It can ensure the mitigation of catastrophe risk within these areas to a greater extent, see Tables  \ref{tab: GRA} and \ref{tab: TOPSIS}. 
\item  The pricing mechanism of a flooding catastrophe bond is studied with considerate trigger choice concerning both the frequency and the layered intensity of the flood disasters via the maximum precipitation. Given the climate changes and financial volatility, the pricing sensitivity of both precipitation severity and financial market distortion convinces the investors and reinsurers of the pricing trend, see Figure \ref{fig: kappa-shape}. 
\end{itemize}

\cL{The paper is organized as follows.} Section \ref{sec: data} shows the  regional and extreme features of  max point precipitation and the caused flooding economic losses. Section \ref{sec: main results} is devoted to the main results %including the extreme analysis of flooding risks, the cross-regional insurance fund and the pricing of the flooding catastrophe bond.
followed by  Section \ref{conclusion} for the conclusions. We end this paper with an appendix in  Section \ref{appendix}, including all technique and methodology involved. %, including EVT modeling, Grey Relational Analysis, Vasicek interest rate modeling and pricing simulation.

\section{Descriptive analysis of flooding catastrophe loss and extreme precipitations}\label{sec: data}
In order to study the catastrophe losses caused by flooding in various regions in China, economic losses of main flooding events as well as its maximum point precipitations (the maximum accumulated precipitation in a particular site within exactly 24 hour) in \CCL{31} provinces of China from 2006 to 2018, which are collected from \href{http://www.mwr.gov.cn/sj/}{http://www.mwr.gov.cn/sj/}, the official website of the Ministry of Water Resources \CCL{of} the \CCL{People's} 
Republic of China. Given the \CCL{money inflation}, all economic loss data is normalized according to 2019 annual consumer price index (CPI). 

We see from Table \ref{tab: descriptive} that,  the average economic loss of the main flooding events in China from 2006 to 2018 is RMB 15.45 billion with a large range from RMB 0.12 \CCL{billion} to RMB 78.65 billion. The economic loss is roughly right-skewed with large kurtosis 5.56, confirmed also by the histogram in Figure \ref{fig: hist-scatter}.  Moreover, the max point precipitation behaves similarly with certain right skewness and deviates slightly from normal distribution as well.  Indeed, the precipitation, one of the most main drivers of floods,  is free of human manipulation. Figure \ref{fig: hist-scatter} shows positive association between economic loss on log scale and max-point precipitations with correlation \CCL{0.45}. This motivates our trigger design of precipitation level for the catastrophe bond in Section \ref{bond}. In particular,  the $70\%, 80\%, 90\%$ and $95\%$ quantile precipitations are 626, 744, 849 and 985, respectively.

\begin{table}[htpb]
\caption{Descriptive analysis of economic loss and maximum point precipitation.}
\centering
\begin{tabular}{cccccccc}
\hline \hline
Variable         & Size & Min& Median & Mean & Max  & Skewness & Kurtosis \\ \hline
Economic loss (billion RMB)               & 94           &  0.12 & 9.69            & 15.45         & 78.65                 & 1.72              & 5.56              \\
Max Point precipitation (mm) & 94            & 75 & 479.5           & 530.07        & 1426                   & 0.79              & 3.65    \\       \hline \hline
\end{tabular}
\label{tab: descriptive}
\\
\hfill{\tiny \textit{source: \href{http://www.mwr.gov.cn/sj/}{http://www.mwr.gov.cn/sj/}, the Ministry of Water Resources of China.}}
\end{table}

\begin{figure}
    \centering
    \includegraphics[width=7cm]{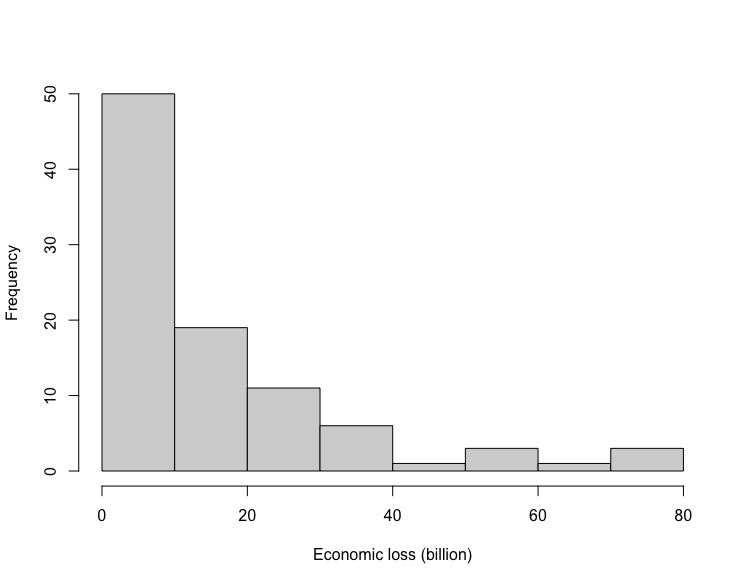}
    \includegraphics[width=7cm]{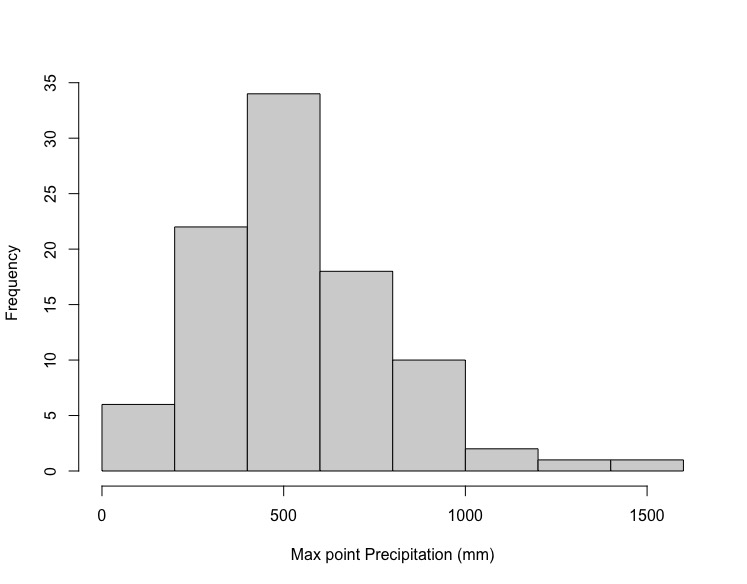}
    \includegraphics[width = 14cm, height=6cm]{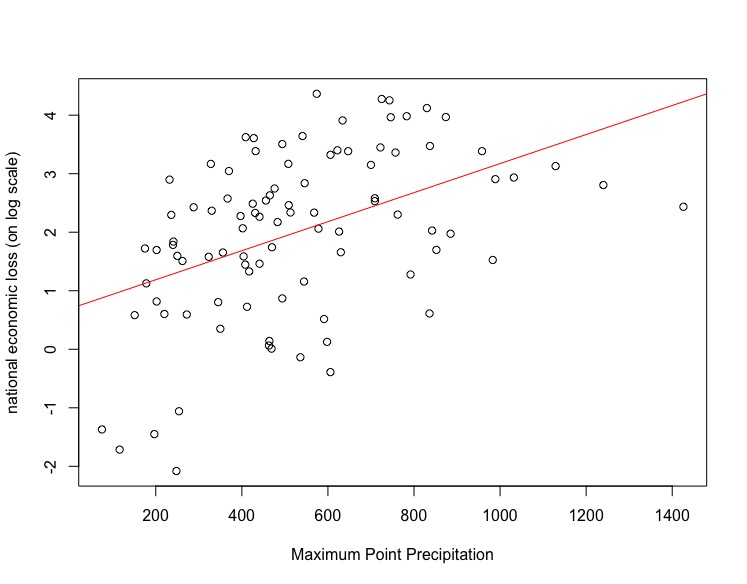}
    \caption{Histogram of economic loss (top left) and maximum point precipitation (top right) and scatter plot of economic loss against max point precipitation (bottom).}
    \label{fig: hist-scatter}
\end{figure}

In order to find the hot-spot regions of flooding risks and propose corresponding regional tackles, we summary the average economic loss of each province in 2006$\sim$2019. It shows a big spatial discrepancy from province to province. We see in Figure \ref{fig: provincial loss} that, the average loss ranges from RMB 2.49 billion of Shanghai to RMB 225.25 billion of Guangdong with average of RMB 
74.29 billion.  The regional difference of flooding risks is roughly reflected by the economic total loss. On the other hand, each province was exposed to timely-varying flooding risk with standard deviation of yearly provincial loss ranging from 3.04 of Ningxia to 204.85 for Hubei. 
Actually, a variety of disaster-inducing factors, risk tolerance and economic development level can systemically affect the flooding-caused loss among different regions. Therefore, the flooding related risk vulnerability level in different zones should be comprehensively considered.

\begin{figure}[htp!]
\centering
    \includegraphics[width=14cm]{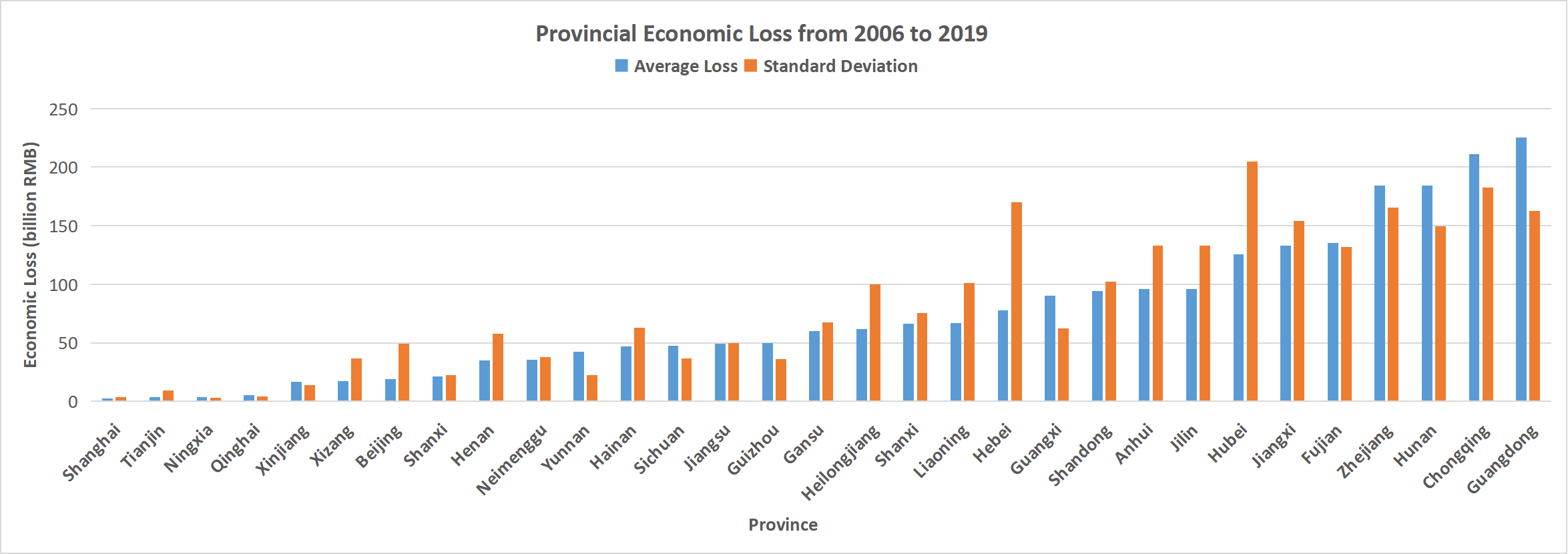}
    \caption{Provincial average economic loss in blue and standard deviation in orange from 2006 to 2019. Data is from \href{http://www.mwr.gov.cn/sj/tjgb/zgshzhgb/}{http://www.mwr.gov.cn/sj/tjgb/zgshzhgb/}, the Ministry of Water Resources of China. }
    \label{fig: provincial loss}
\end{figure}

This motivates our study on cross-regional insurance fund using multiple-criterion decision-making method in Section \ref{cross-regional}. For simplicity, we focus on the flooding vulnerability in 2019 and consider 19 provinces  as representatives of different economic loss levels in Figure \ref{fig: provincial loss},  which are located in the seven primary geographical divisions of China: Northeast, North, Northwest, Southwest, South, East and Central China, including Shanghai, Jiangxi, Zhejiang, Sichuan etc.  

Besides the direct economic loss, another 12 indexes listed in Table \ref{tab: indicator}  below  are also selected as the evaluation criteria for flooding disasters, among which the vulnerability level of flooding disasters can be mainly divided into two parts: risk exposure and regional economic development level, which are from   \emph{the Bulletin of Flood and Drought Disasters in China} and \emph{the Provincial Yearbook 2019}, respectively.   

To summary, the economic loss caused by severe floods behaves with certain regional difference and highly correlated relationship with extreme precipitations. In the following section, we will establish extreme models to conduct the flooding risk management through the insurance, reinsurance and financial markets.

\begin{table}% {\linewidth}
\captionof{table}{Flood catastrophe vulnerability evaluation indicators}
	\centering
\small
\resizebox{0.9\textwidth}{!}{
\begin{tabular}{l|l|l|l}
\hline\hline
Group                                                & %\multicolumn{1}{c}{Indicator} &       
{Label} & {Description} & Units              \\
\hline
\multirow{6}{*}{{Risk Exposure}}                                                                        & $X_1$                            & Suffered Population                                  & 10000 persons      \\
& $X_2$                            & Disastered Farmlands                                 & 1000 Ha            \\
 & $X_3$                            & Collapsed Houses                                     & million rooms      \\
& $X_4$                            & Number of Damaged Dams                               & one                \\
& $X_5$                            & Length of Damaged Dams                               & km                 \\
 & $X_6$                            & Direct Economic Losses                               & 100 million RMB       \\
 \hline
\multirow{7}{3cm}{{Regional Economic Development Level}} & $X_7$ (-)\tnote{1}                       & Per Capital GDP                                      & yuan               \\
                                                     & $X_8$ (-)                        & Per Capital Urban Disposable Income                  & yuan               \\
                                                     & $X_9$ (-)                        & Local Financial Revenue                              & 100 million yuan   \\
                                                     & $X_{10}$ (-)                       & Number of Sanitary Beds                              & Sets/10000 persons \\
                                                     & $X_{11}$ (-)                       & Number of Graduates from Higher Education Instiution & one                \\
                                                     & $X_{12}$ (-)                       & Insurance Density                                    & \%                 \\
                                                     & $X_{13}$ (-)                       & Insurance Penetration                                & \%   \\
                                                     \hline\hline
\end{tabular}
}
\par
	\begin{tablenotes}
	\footnotesize
	\item[1] (-) sign represents the negative relationship between the indicator and flooding risk vulnerability. The data is available in {the Bulletin of Flood and Drought Disasters in China} and {the Provincial Yearbook 2019}. 
	\end{tablenotes}
\label{tab: indicator}
\end{table}

\section{Main Results}\label{sec: main results}
In risk management, of importance is to design a layered compensation among all stakeholders and diversify potential risk in insurance, reinsurance and financial markets. This section will give the main results concerning the extreme models of flooding economic losses and precipitations in Section \ref{sec: EVA}, which will be applied to determine layered compensation scheme as well as the natural disaster risk in the price of flooding catastrophe bond in Section \ref{cross-regional} and \ref{bond}, respectively. 

\subsection{Extreme analysis of flooding economic losses and precipitations}\label{sec: EVA}
In this section, we apply Extreme Value Theory (EVT) to analyse the tail distribution law of the economic loss caused by severe floods. Given the limited datasets, we consider the compound Poisson Process of excess loss over a high threshold with severity following generalized Pareto distribution and Poisson frequency distribution. More specifically, let $S_{N}$ be the total excess loss during 2006-2018, given by 
$$S_N = \sum_{i=1}^N X_i,$$
where $X_i$'s are the excess loss before Year $t$ and $N=N(t)\sim Poisson(\lambda t)$ is the number of severe floods or excess losses before Year $t$. Suppose that $N$ and $X_i$'s are independent. Thus, the expected total loss is obtained by (cf. \eqref{eq: mep})
\BQN\label{eq: total loss}
\E{S_N} = \E{N}\E{X} = \lambda \left(\frac{u}{1-\xi}+\frac{\sigma - \xi \mu}{1-\xi}\right),
\EQN
where $\lambda$ is the mean occurrence of extreme economic loss over certain threshold $u$ specified below, and $(\mu, \sigma, \xi)$ is the triplet of parameter involved in the generalized extreme value distribution (GEV) in \eqref{GEV}. 
In the following, we employ the mean residual life plot and variation plot to determine the economic loss threshold which helps to identify the GP distributed excess losses $X_i$'s. 

\CCL{We see from} Figure \ref{fig: mrp-variation} that the mean residual life plot tends to be linear in threshold within 15 to around 30. The stability of parameter estimates across the reduced threshold range 15 to 30 are demonstrated \CCL{in the variation plots}. Short bars are relatively better choices. Among them, a lower value of the threshold is suggested by \citep{Coles2001} in order to ensure enough exceedances for accurate maximum likelihood estimate (MLE). As a result, threshold $u$ is set as \CCL{17.19, the corresponding $70\%$ empirical quantile.} 

\begin{figure}[htpb]
    \centering
    \includegraphics[width=8cm]{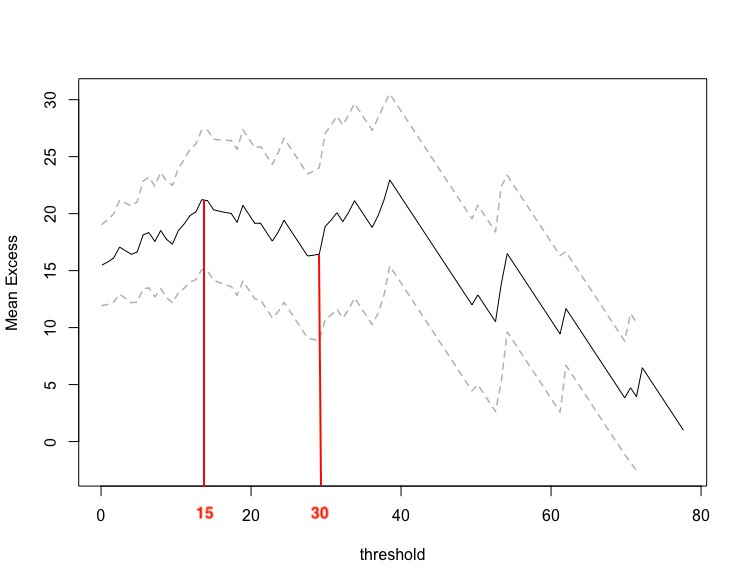}
\includegraphics[width=8cm]{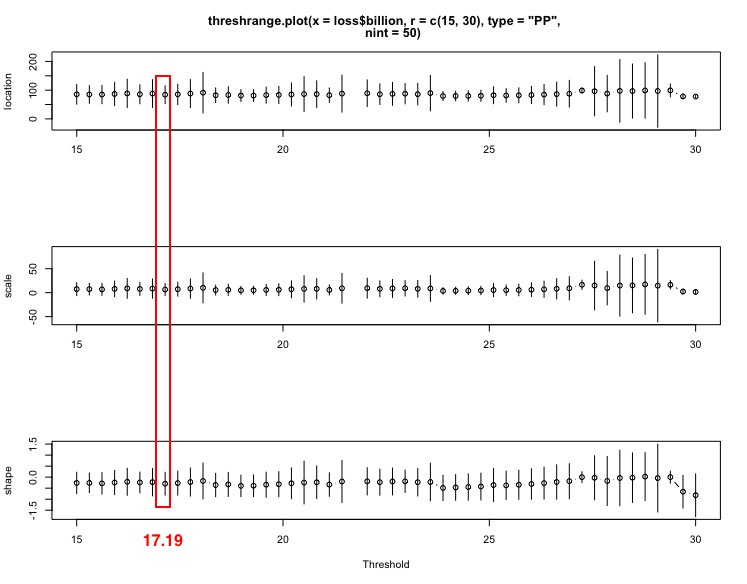}
    \caption{Mean residual life plot of economic loss (in billion yuan) with grey dash line representing 95\% confidence interval (left). \cL{Variation plots of} location, scale, and shape (from above to bottom) against threshold $u\in(15, 30)$ (right).  
    }
    \label{fig: mrp-variation}
\end{figure}

Now, we apply the excess models given in Section \ref{app: EVT}, namely we fit the excess losses by Point Process, compared with GP model and exponential model (the reduced model of GP with $\xi=0$) as well. The obtained estimates of the parameters are shown in Table \ref{tab: PP}. We see that Point Process  model with minimum AIC and BIC is a challenging model to fit the excess losses. We get the maximum-likelihood estimate of the location, scale, and shape parameters of $(\mu,\sigma,\xi)$ is obtained by maximizing \eqref{eq: loglik} as follows.
$$ (\widehat\mu,\widehat\sigma,\widehat\xi) = (83.34, 6.64, -0.29).$$
Similar procedure applies in the threshold excess of maximum point precipitation and gives the threshold of 626 and parameter estimated given in Table \ref{tab: vasicek} in Section \ref{bond}. 

We also carry out the log-likelihood ratio test between PP model and GP (exponential) model. It turns out that PP model overwhelms the other two models with $p\leq2.2\times 10^{-16}$. Diagnostic plots of the PP model and GP model for threshold economic loss and precipitation are displayed in  Figure \ref{fig: qq}. Points in these two figures can be generally considered as close to the unit diagonal, which means that exceedances are well fitted to the presupposed model.

\begin{table}[htpb]
\caption{AIC, BIC and estimates of location, scale and shape parameters with standard errors in parentheses when fitting economic excess losses by Point Process, Generalized Pareto and Exponential models. Here threshold $u = 17.19$.}
\resizebox{\textwidth}{!}{
\begin{tabular}{cccccccccccc}\hline \hline
\multirow{2}{*}{Model} & \multicolumn{2}{c}{Location parameter ($\mu$)} & \multicolumn{2}{c}{Scale parameter ($\sigma$)} & \multicolumn{2}{c}{Shape parameter ($\xi$)} & \multirow{2}{*}{AIC} & \multirow{2}{*}{BIC} \\ \cmidrule(r){2-3} \cmidrule(r){4-5} \cmidrule(r){6-7}
                       &Estimates (s.e.)  & 95\% CI & Estimates (s.e.)   & 95\% CI   & Estimates (s.e.) & 95\% CI &                      &                      \\ \hline
Point Process    & 83.34 (16.11)          & (52.77, 115.91)           & \ 6.64 (6.44)             & $(-5.99, 19.27)$        & $-$0.29 (0.27)            & ($-$0.81, 0.23)     &\  22.16                &\  26.16                \\
Generalized Pareto     & -                    & -                        & 27.11 (7.96)             & (11.50, 42.71)        & $-$0.32 (0.23)            & ($-$0.78, 0.13)     & 226.76               & 229.42               \\
Exponential            & -                 & -                        & 20.18 (3.81)             & (12.71, 27.66)        & - & -                                 & 226.27               & 227.60 
\\ \hline \hline
\end{tabular}
}
\label{tab: PP}
\end{table}

\begin{figure}
    \centering
    \includegraphics[width=7cm]{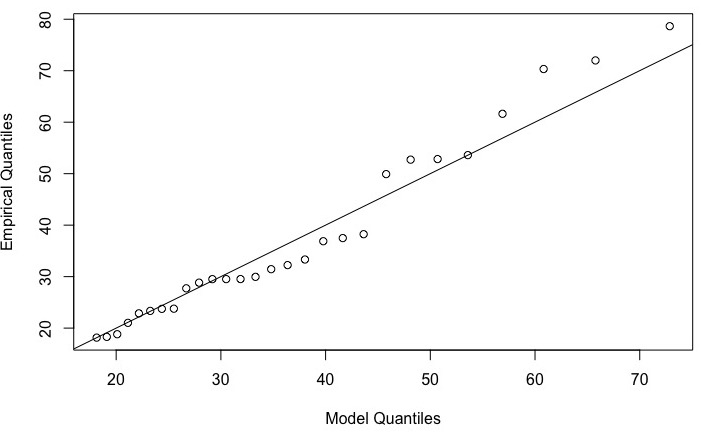}
    \includegraphics[width=7cm]{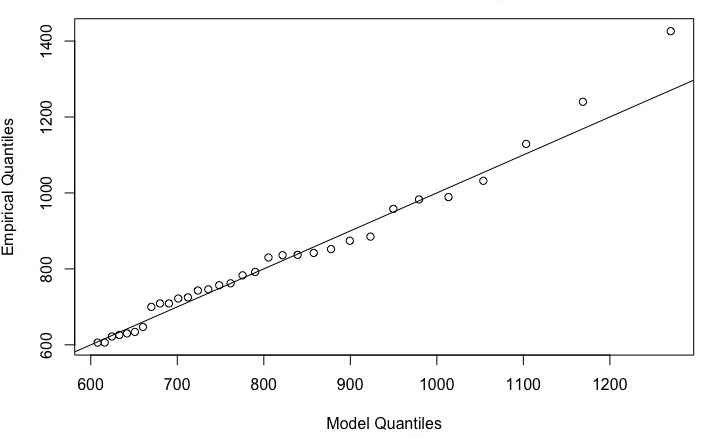}
    \caption{QQ plot of economic loss (left) and maximum point precipitation (right).}
    \label{fig: qq}
\end{figure}

To get the expected total annual loss, it remains to check if the frequency of excess loss follows Poisson distribution. We employ  Kolmognov-Smirnov test and obtain $p$-value equal $0.3626$, which suggests a good Poisson fit resulting in $\widehat \lambda= 2.55$. Consequently, the total annual economic loss is estimated as \CCL{RMB $2.55\times 33.84 = 86.29$ billion}. 

Note that VaR and CVaR are two common risk measures applied to evaluate the extreme risks with good mathematical properties in insurance and finance fields. Applying extrapolation approach (c.f. \eqref{eq: VaR} and \eqref{eq: CVaR}), Table \ref{tab: VaR-CVaR} shows the estimates of VaR and CVaR values at different confidence levels formalizing the compensation system. Its back-testing shows the efficiency of all estimates in Table \ref{tab: backtesting}. For both VaR and CVaR models, %they cover the expected economic loss RMB 33.84 billion for all confidence levels, and 
its spillover rate as a degree of risk coverage is fairly close to the significance value, implying the accurate estimate. 

 In particular, we specify four loss levels of compensation mechanism using the VaR and CVaR values, which can be covered among all stakeholders, including the insurers, local government, reinsurance and government fund like catastrophe bond. Besides the small parts of loss level (less that $85\%$ VaR) covered by the insurers, we consider a zoning insurance fund for the moderate insurance cover (between $85\%$ VaR and $90\%$ CVaR) in Section \ref{cross-regional}. While for the higher loss level,  the insurance companies may generally transfer the larger risk to reinsurance as well as the financial markets in terms of various financial securities (\citep{Zhou2017, tang_yuan_2019}). Thus, in Section \ref{bond} we investigate the price of flooding catastrophe bond.

\begin{table}[htpb]
    \caption{Design of compensation mechanism.} 
    \begin{threeparttable}
    \centering
    \resizebox{0.9\textwidth}{!}{
    \begin{tabular}{cccccc}
    \hline \hline
   Confidence level & VaR    & CVaR   & Loss level   & Loss amount (billion) & Loss taker                                         \\ \hline
    97.5\%           & 58.26 & 75.77 & Fourth level  & \textgreater{}\ 64.70     & Government (CAT bond)                                            \\
    95.0\%             & 46.97 & 64.70 & Third level & 53.48--64.70      &  Re-insurance             \\
    90.0\%             & 35.53 & 53.48 & Second level  &
    28.76--53.48          & CRFCIF*                                       \\
    85.0\%             & 28.76 & 46.85 & First level & \textless{}\ 28.76  & Insurer                                                 \\ \hline \hline
    \end{tabular}
    }
\begin{tablenotes}
      \small
      \item CRFCIF*: Cross-Regional Flooding Catastrophe Insurance Fund.
\end{tablenotes}
\end{threeparttable}
     \label{tab: VaR-CVaR}
\end{table}

\begin{table}[htpb]
\caption{Value and spillover rate of VaR and CVaR.}
\resizebox{\textwidth}{!}{
\begin{tabular}{ccccccc}
\hline \hline
\multirow{2}{*}{Significance value} & \multicolumn{3}{c}{VaR}                                    & \multicolumn{3}{c}{CVaR}                                   \\ \cmidrule(r){2-4} \cmidrule(r){5-7}
                                    & Value & Theoretical spillover (\%) & Actual spillover (\%) & Value & Theoretical spillover (\%) & Actual spillover (\%) \\ \hline
1.0\%                                 & 72.94 & 1(1.0\%)                     & 1(1.06\%)             & 90.16 & 1(1.0\%)                     & 0(0.00\%)                \\
2.5\%                               & 58.26 & 2(2.5\%)                   & 4(4.26\%)             & 75.77 & 2(2.5\%)                   & 1(1.06\%)             \\
5.0\%                                 & 46.97 & 5(5.0\%)                     & 8(8.51\%)             & 64.70 & 5(5.0\%)                     & 3(3.19\%)             \\
10\%                                & 35.53 & 9(10\%)                    & 11(11.70\%)           & 53.48 & 9(10\%)                    & 5(5.32\%)             \\
15\%                                & 28.76 & 14(15\%)                   & 19(20.21\%)           & 46.85 & 14(15\%)                   & 8(8.51\%)   \\ \hline \hline         
\end{tabular}
\label{tab: backtesting}
}
\end{table}

\subsection{Design of Cross Regional Compensation}\label{cross-regional}

This section is devoted to the design of zoning insurance level according to its flooding vulnerability. 
We apply two common approaches in terms of Multiple-Criteria Decision-Making in Appendix \ref{app: MCDM}. Comparison analysis of the rank of the 19 flood-prone provinces is conducted according to the selection of risk indicators (cf. Table \ref{tab: indicator}) and ranking indices given by Grey Relational Analysis (GRA) and the Technique for Order Preference by Similarity to an Ideal Solution (TOPSIS), see  Appendices \ref{app: GRA} and \ref{app: TOPSIS}.

\subsubsection{Ranking of flooding vulnerability via Grey relational analysis (GRA)}

Grey relational analysis  (GRA) is one of the most significant multivariate statistical analysis methods in the field of Grey System, which is mainly used to evaluate the relative performance or optimize multi-parameters to obtain the best quality characteristics among different discrete sequence 
(\citep{Deng1982}, \citep{Zhao2013}, \citep{Peng2021ranking}). Given the original small sample size of the rare extreme flooding events involved, it seems that the GRA may be the best method to do this study since it is commonly used in dealing with problems of sample size and distribution uncertainty (\citep{Ho2006}). 
Hence, we shall primarily implement the grey relational analysis  method to evaluate and make a classification to the vulnerability level (premium level) for 19 provinces in China, based on the indicators relevant with flooding risks in 2019 (see Table \ref{tab: indicator} in Section \ref{sec: data} above). This will be helpful for future insurance premium allocation.

Based on the algorithm of grey relational grade calculation in Appendix \ref{app: GRA}, Table \ref{tab: GRA} shows the flooding risk ranking of 19 provinces and compares the rank if we select risk exposure indexes and exclude (include) the consideration of  economic development level in different regions (see also Figure \ref{fig: F1}).
 \begin{table}[htb!]
\caption{Rank of {each province in flooding vulnerability based on} all indicators (left) and without regional economic indicators (right). Equal weight applies in the grey relational analysis.}
    \resizebox{0.8\textwidth}{!}{
    \begin{minipage}{\linewidth}
\begin{tabular}{rcr}
\hline\hline
{Province} 
& Grey Relational Grade A               & Ranking               \\
 \hline
\LE{Jiangxi    }               & \LE{0.6882 }                             & \LE{1}                      \\
Heilongjiang              & 0.6867                              & 2                           \\
Hunan                     & 0.6468                              & 3                           \\
Yunnan                     & 0.6027                              & 4                           \\
\hline
Gansu                     & 0.5846                              & 5                           \\
Shandong                  & 0.5823                              & 6                           \\
Anhui                     & 0.5792                              & 7                           \\
Sichuan                   & 0.5687                              & 8                           \\
Shanxi                    & 0.5605                              & 9                           \\
Jilin                     & 0.5535                              & 10                          \\
Inner Mongolia                 & 0.5431                              & 11                          \\
Hubei                     & 0.5154                              & 12                          \\
Chongqing                 & 0.5149                              & 13                          \\
Zhejiang                  & 0.5109                              & 14                          \\
\hline
Henan                     & 0.4953                              & 15                          \\
\LP{Shanghai}                  & \LP{0.4390}                             & \LP{16}                          \\
Guangdong                 & 0.4344                              & 17                          \\
Beijing                   & 0.4207                              & 18                          \\
Jiangsu                   & 0.4118                              & 19            \\
\hline\hline
\end{tabular}
\hspace{0.1cm}
\begin{tabular}{rcr}
\hline\hline
{Province} &  Grey Relational Grade B                & Ranking                \\
 \hline
\LE{Jiangxi  }                 & \LE{0.6299  }                              & \LE{1}                            \\
Heilongjiang              & 0.6234                                & 2                            \\
Shandong                  & 0.5994                                & 3                            \\
Hunan                     & 0.5679                                & 4                            \\
\hline
Zhejiang                  & 0.4935                                & 5                            \\
Sichuan                   & 0.4780                                & 6                            \\
\hline
Guangdong                 & 0.3826                                & 7                            \\
Shanxi                    & 0.3624                                & 8                            \\
Hubei                     & 0.3611                                & 9                            \\
Jilin                     & 0.3591                                & 10                           \\
Gansu                     & 0.3494                                & 11                           \\
Yunnan                     & 0.3484                                & 12                           \\
Chongqing                 & 0.3449                                & 13                           \\
Anhui                     & 0.3441                                & 14                           \\
Henan                     & 0.3414                                & 15                           \\
Inner Mongolia                 & 0.3396                                & 16                           \\
Jiangsu                   & 0.3361                                & 17                           \\
Beijing                   & 0.3334                                & 18                           \\
\LP{Shanghai}                  & \LP{0.3334}                               & \LP{19}       \\
\hline\hline
\end{tabular}
\end{minipage}
}
\label{tab: GRA}
\end{table}
\par
}

We see that three zoning categories A, B and C with high, moderate and low premium payments can be determined according to the decreasing flood vulnerability of each province. Compared with the results including regional economic development indicators, it is can be found that ranking location for many provinces has changed. This seems rational in view of \citep{Zhou2017} concerning the criterion of ‘either the area with serious loss or the developed area undertakes more’ in their study on Typhoon catastrophe. That means for some underdeveloped regions, the appropriate privilege can be undertaken for the insurance premium. For instance, Yunnan,  Inner Mongolia and Gangsu drop while Shandong, Zhejiang and Jiangsu rises to some extent. The change of ranking order can be largely interpreted by the economic development situation. 
In terms of descending order in some provinces, when collecting data, the economic development criterion of these provinces like per capita GDP, per capita disposable income and local fiscal revenue are found to be relatively lower than other provinces. Hence, it is significant to redistribute the initial capital of the flooding catastrophe fund to relieve the resource and financial burden in some less developed regions. And it is also obvious that the ascending ranking are often occurred in the coastal provinces like Shandong, Zhejiang, Shandong. This reflects these provinces may have more complete social security, more solid financial capacity, more cutting-edged medical facilities, and stronger risk prevention and protection awareness in supporting them to undertake possible flooding-related loss.

\begin{figure}[htp!]
\centering
    \includegraphics[width=15cm]{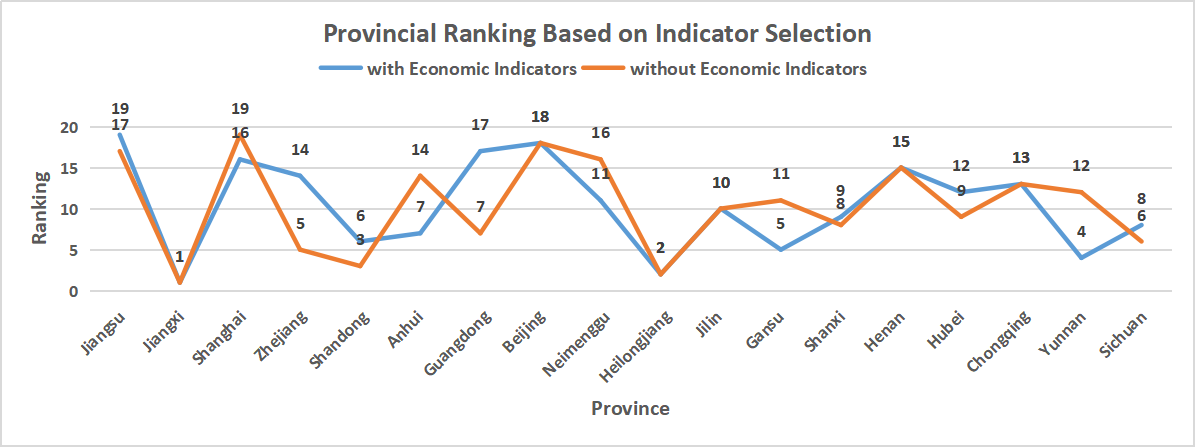}
    \caption{Provincial Ranking Based on Indicator Selection.}
    \label{fig: F1}
\end{figure}

\subsubsection{The Technique of Order Preference by Similarity to an Ideal Solution (TOPSIS)}
In comparison of the result of grey relational analysis,  this section is dedicated to the  entropy weight method and The Technique for Order Preference by Similarity to an Ideal Solution (TOPSIS)  based on closeness coefficient (\citep{Sun2020}), see Section \ref{app: TOPSIS}.
Besides, the TOPSIS  effectively reduces the influence from the subject choice of distinguishing coefficient involved in the GRA. Here, all 13 flooding risk indicators are included and the entropy weights apply in both GRA and TOPSIS. 

\begin{figure}[htp!]
\centering
    \includegraphics[width=15cm]{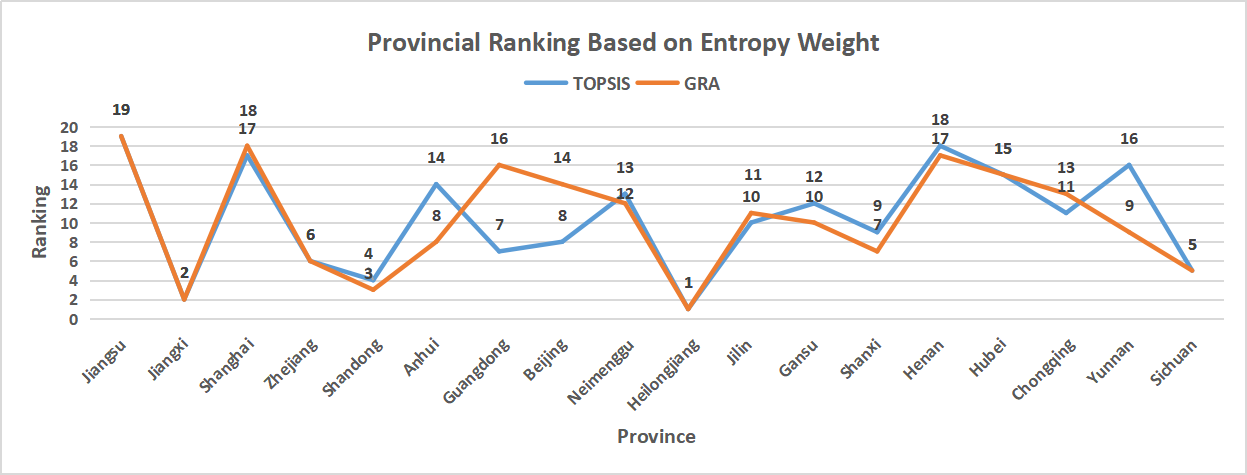}
    \caption{Provincial Ranking Based on Entropy Weight.}
    \label{fig: F2}
\end{figure}

Table \ref{tab: TOPSIS} shows the rank of flooding vulnerability according to 
the calculated closeness coefficient and relational grade involved in TOPSIS and GRA. 
Note that the province with larger closeness coefficient implies the more severe flooding damage it may suffer from since it decreases in the relative distance to the highest 
risk vulnerability province compared to the lowest one, see Eq. \eqref{CC}. We see most provinces (11 out of 19) get rather lower closeness coefficients, which are thus classified with low insurance premium level. On the other hand, most provinces remain steady ranking under GRA and TOPSIS except Anhui, Yunnan, Beijing and Guangdong (see Figure \ref{fig: F2}). There are several reasons accounting for this change. Different normalization method in GRA and TOPSIS method may explain this change. GRA method uses the linear normalization (also called min-max normalization), where the maximum and the minimum value of each feature of the dataset are used to scale the value of the feature to the interval [0,1], achieving the equal scaling of the original data (Eq. \eqref{normalizationGRA}). While in TOPSIS, the minimum and maximum value are taken to divide an to be divided by original data respectively, to form a new sample dataset (Eq. \eqref{normalizationTOPSIS}). Besides, from the entropy weight calculation results, higher weights are assigned to the original risk exposure indicators ($X_1\sim X_6$) , while weights of regional economic indicators are reduced to a lower level.

\begin{table}[htb!]
\caption{Rank of each province in flooding vulnerability based on closeness coefficient (left) and GRA (right).  The shannon entropy weights apply in both methods with all indicators taken into account.}
    \resizebox{0.8\textwidth}{!}{
    \begin{minipage}{\linewidth}
    \begin{tabular}{rcr}
    \hline\hline
    Province     & Closeness Coefficient & Ranking \\ \hline
    Heilongjiang & 0.5471                & 1       \\
    Jiangxi      & 0.5180                & 2       \\
    Hunan        & 0.4879                & 3       \\
    Shandong     & 0.4819                & 4       \\
    \hline
    Sichuan      & 0.3758                & 5       \\
    Zhejiang     & 0.3288                & 6       \\
   \hline
    Guangdong    & 0.1918                & 7       \\
    Beijing      & 0.1914                & 8       \\
    Shanxi       & 0.1456                & 9       \\
    Jilin        & 0.1282                & 10      \\
    Chongqing    & 0.1279                & 11      \\
    Gansu        & 0.1223                & 12      \\
    Inner Mongolia    & 0.1187                & 13      \\
    Anhui        & 0.1156                & 14      \\
    Hubei        & 0.1151                & 15      \\
    Yunnan        & 0.1097                & 16      \\
    Shanghai     & 0.0975                & 17      \\
    Henan        & 0.0654                & 18      \\
    Jiangsu      & 0.0300                & 19     \\ \hline\hline
    \end{tabular}
    \hspace{0.1cm}
    \begin{tabular}{rcr}
    \hline\hline
Province     & Grey Relational Grade &  Ranking \\ \hline
Heilongjiang & 0.6422                & 1             \\
Jiangxi      & 0.6124                & 2             \\
Shandong     & 0.5782                & 3             \\
Hunan        & 0.5711                & 4             \\
\hline
Sichuan      & 0.4902                & 5             \\
Zhejiang     & 0.4729                & 6             \\
Shanxi       & 0.4371                & 7             \\
Anhui        & 0.4308                & 8             \\
Yunnan       & 0.4269                & 9             \\
Gansu        & 0.4200                & 10            \\
Jilin        & 0.4153                & 11            \\
Inner Mongolia    & 0.4081                & 12            \\
Chongqing    & 0.4040                & 13            \\
Beijing      & 0.4005                & 14            \\
\hline
Hubei        & 0.3977                & 15            \\
Guangdong    & 0.3970                & 16            \\
Henan        & 0.3801                & 17            \\
Shanghai     & 0.3787                & 18            \\
Jiangsu      & 0.3572                & 19            \\ \hline\hline
     \end{tabular}
     \end{minipage}
    }
    \label{tab: TOPSIS}
    \end{table}

\subsection{Design of Flooding Catastrophic Bonds}\label{bond}

\begin{figure}[htpb]
    \centering
    \includegraphics[scale = 0.4]{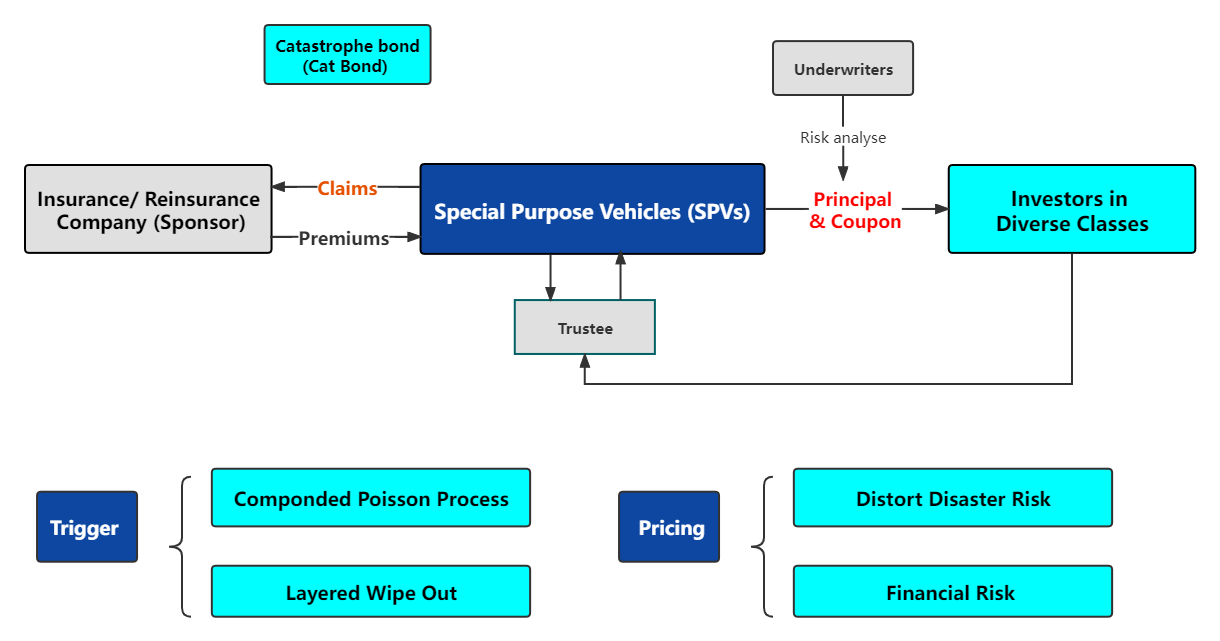}
    \caption{Operation mechanism for catastrophe bond.}
    \label{workflow for bond}
\end{figure}

Because of the low frequency and high severity features of flooding events, insurers and re-insurers are eager to hedge their risk. To this, the bond market is a sounding choice since investors looking for arbitrage opportunities and finally mitigate the loss of insurance products. Collateralize special purpose vehicles (SPVs) issue the CAT bond, which is usually issued and established by sponsors who are insurers and re-insurers. For the SPVs, the premium is received from the sponsors and reinsurance coverage are given in return. The rewarded premium usually is paid to the investor as a part of coupon payment and it contains a floating portion related to national reference rate. For example, London Interbank Offered Rate (LIBOR), Shanghai Inter-bank Offered Rate (Shior), can reflect the return from the trust account where the principal is deposited. The principal and coupon payments will be reduced whenever a specific triggering event happens. Also, some funds can be sent to the sponsor as a reimbursement for the claims. \CCL{Here we follow the product pricing scheme proposed by \citep{tang_yuan_2019} together with a compound Poisson trigger process similar to the 2015 Acorn Re earthquake CAT bond.} The detailed workflow is presented in Figure \ref{workflow for bond} .

Here, we consider a payoff function $\Pi (\cdot )$ and wipe out time $\tau$ defined by
\begin{equation*}
\Pi (y)= \max\left \{ 1-y,0 \right \} ,\ y\ge 0,\quad \tau = \inf\{t\ge0: \Pi(Y_t)=0\}.
\end{equation*}
Here the trigger process  $\{Y_t,\ t\ge0\}$ is supposed as a compound Poisson trigger process  given below
\BQNY
Y_{t}  &=&\sum_{j=1}^{N_{t} }  (0.005\times \I{X_{j}\in (626,744] }+ 
0.015\times \I{X_{j}\in (744,849] } \nonumber\\
&&\quad+ 0.15\times \I{X_{j}\in (849,985] }+ 
0.2\times \I{X_{j}\in (985,\infty] } ),
\EQNY
where $X_{j}$ denotes the severity of flooding driver (here max-point precipitation) and $N_{t}$ denote the number of precipitation excess by time $t$. These four precipitation levels of 626, 744, 849 and 985 account for 0.7, 0.8, 0.9 and 0.95 quantile accordingly,  which means when severity above certain level of threshold, a fraction $0.5\%, 1.5\%, 15\%, 20\%$ of  \CCL{principal will be wiped off.}

Hencee, for  a $T=3$ year CAT bond with face value $K=1000$, in which each coupon period equals $\Delta =1/4$, and hence coupons should be paid at times $s\Delta$, where $s=1,\ldots, 4T$. 
%Then the accrued coupon when the bond is terminated before maturity becomes $K\vartheta$ with 
%\begin{equation*}\vartheta = (\tau  - \left \lfloor 4\tau   \right \rfloor )(R + i_{\tau })\CCL{\Pi(Y_{\lfloor \tau \rfloor})}.
%\end{equation*}
And it follows from Eq. (\ref{price}), the  price at time $t$ becomes
\begin{align}\label{price0}
P_{0}=& \frac{K}{4} \EE^{Q^{1}}\left[\sum_{s=1}^{\lfloor 4 \tau\rfloor \wedge 4 T} \EE_{t}^{Q^{2}}\left[D(0, s \Delta)\left(R+i_{s \Delta}\right)\CCL{\Pi(Y_{(s-1)\Delta})}\right]\right]+K Q^{1}(\tau>T) \EE^{Q^{2}}[D(0, T)\Pi(Y_T)]. \\ \nonumber
&+\frac{K}{4} \EE^{Q^{1}}\left[(\tau-\lfloor 4 \tau\rfloor \Delta) \I{\tau \leq T}\CCL{\Pi(Y_{\lfloor \tau \rfloor})} \EE^{Q^{2}}\left[D(0, \tau)\left(R+i_{\tau}\right) \mid \tau\right]\right]. \nonumber
\end{align}
Note that pricing scheme can only be realized by Monte-Carlo simulation. To this, we model the precipitation excess by generalized Pareto distribution $GP_{\xi, \beta}$. 
The threshold of $u=600$ is determined by the mean residual plot and parameter variation plots, similar to that for excess economic loss in Figure \ref{fig: mrp-variation}. Meanwhile, we fit the annual occurrence process $\{N(t),\ t\ge0\}$ of precipitation excess with Poisson distribution and get the intensity  $\lambda $ equals to 2.55. Furthermore, we implement a Wang's distortion of the disaster risk $X$ (\citep{wang2000class}), i.e., 
$$\widetilde X =\left [ (1-\Phi(\Phi^{-1}(U )+\kappa))^{-\xi } -1 \right ] \times  \frac{\beta}{\xi } + u.$$
Additionally, the financial interest risk $(r_t, \ell_t)$ in Eq. \eqref{eq: financial risk} is fitted by the 4 year 3-month China treasury bond rates and 4 year 3-month Shanghai Inter-bank Offered Rate (Shibor). Estimated parameters are put into the pricing measure that combines a distorted pricing measure represent catastrophe insurance risk and risk-neutral pricing measure for the arbitrage financial market. The bond price is found by a simulation of $10^5$ paths each with \CCL{distorted generalized Pareto distributed inter-arrival time (severity of max-point precipitation excess)}.

\begin{table}[htpb]
\centering
\caption{Specification of interest rate processes and precipitation distribution.}
\begin{tabular}{cccccccccc}
\hline \hline
\multicolumn{9}{c}{Panel A: Vasicek models (under $Q^{2}$)}                                      \\ \hline 
\multicolumn{4}{c}{Risk-free rate}             & \multicolumn{4}{c}{Shibor}         & Correlation \\ \cmidrule(r){1-4} \cmidrule(r){5-8}  \cmidrule(r){9-9}
 $a_{r}$    &  $b_{r}$        & $\sigma_{r} $    & $r_{0}$        & $a_{\ell}$       & $b_{\ell}$       & $\sigma_{\ell} $   & $\ell _{0}$     & $\rho $          \\ 
 1.52       &  4.12\%    &  1.40\%    & 2.28\%   & 0.04     & 2.02\% & 4.00\% & 2.43\% & 0.89       \\  \hline
\multicolumn{9}{c}{Panel B: Maximum point precipitation distribution}                                          \\ \hline
\multicolumn{4}{c}{Under $\mathbb P^1$}                   & \multicolumn{5}{c}{Under $ Q^{1}$}                    \\ \cmidrule(r){1-4} \cmidrule(r){5-9}
\multicolumn{4}{r}{$\pk{X >u}=34\%$}& \multicolumn{5}{c}{$Q^{1}  (X \le x)=\Phi (\Phi ^{-1}  (P^{1}  (X \le x)) - \kappa )$}                        \\
\multicolumn{4}{r}{$X- u | (X > u) \sim  GPD _{\gamma,\beta}$ } & \multicolumn{5}{c}{$\kappa=0.42$}                      \\
\multicolumn{4}{r}{$\gamma = -0.181, \beta = 258.55$}           & \multicolumn{5}{c}{\multirow{2}{*}{}}           \\  
\multicolumn{4}{r}{$u=600$}   \\
\multicolumn{4}{r}{Maximum magnitude: 2032.31}  
\\\hline \hline
& \multicolumn{5}{c}{}
\label{tab: vasicek}
\end{tabular}
\end{table}

\begin{figure}[htpb]
    \centering
    \includegraphics[width=7cm]{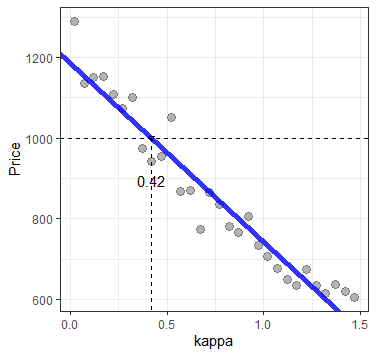}
    \includegraphics[width=7cm]{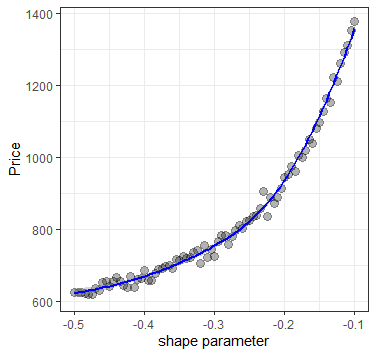}
    \caption{\cL{The price at $t=0$ varies in the distortion parameter $\kappa$ (left) and  in the shape parameter $\xi$ (right). Here, we simulate $10^5$ samples of the threshold excess of point precipitation from $GPD_{\xi, \beta}$. The other parameters specified in Table \ref{tab: vasicek} for the left plot, and we keep $\beta/\xi = -1432.311, \kappa = 0.42$ for the right plot.} 
    }
   \label{fig: kappa-shape}
\end{figure}

Figure \ref{fig: kappa-shape} shows the price varies in the financial market risk in term of the distortion parameter $\kappa\in(0,1.5)$ of Wang's transform. A larger value of $\kappa$ means a higher risk that investors are exposed to, so a higher premium they will require, and eventually a lower bond price. In addition,  the practical value of $\kappa = 0.41$ is determined when it allows par value (1000) to be equal to the bond price. Furthermore, the sensitivity test is conducted for its shape parameter, presenting in \cL{Figure \ref{fig: kappa-shape}}. At this time, the ratio of scale $(\beta)$ to shape $(\xi)$ remains to be $-1432.311$, which means the maximum (the right endpoint) is kept unchanged as 2032.31. 
It turns out that, the bond price increases in negative shape parameter, which is reasonable in reality as investors are exposed to higher risk and require lower cost/ higher return at this time.

\section{Conclusions}\label{conclusion} 

This paper addresses the layered compensation scheme of flood-caused economic loss and the cross-regional insurance fund in the framework of extreme value theory and multiple-criteria decision-making. 
Besides, to mitigate the risk among insures and reinsurers,  further study on flood catastrophe bond price is conducted to reflect both the financial market risks and potential disaster risk with the primary flood driver of extreme precipitations as triggers. 
We provide the government and the local flooding disaster risk management with the catastrophic funding operation mechanism. Moreover, the risk management policymakers should consider regional difference among the disaster risk exposures, the economic development level and the local living conditions.  The financial securities including natural disaster bonds can relieve the traditional insurance companies’ pressure and disperse the possible huge loss into both insurance company and capital markets, which can return attract more investors to participate in the risk dispersion and transfer of flooding catastrophes.

\section{Appendix}\label{appendix}
\subsection{Extreme Value Theory}\label{app: EVT}
Extreme value theory is a natural tool to model severe floods and its caused economic losses since few extreme precipitations results in more than 90\% annual economic losses. The flood risk management institutes should focus on the maxima and tail exposures instead of expected average outcome of compartmental models. 

Given a high threshold $u$, suppose that the threshold excess of $X$ is well fitted by a Generalized Pareto (GP) distribution $G(y; \sigma(u),\xi)$. Then the exceedance probability of $X$ over larger $x$,  can be represented as
\begin{eqnarray}\label{Fx}
    1-F(x)=(1-F(u))\pk{X>x|X>u}  \approx  (1-F(u))\left(1+\xi\frac{x-u}{\widetilde\sigma}\right)^{-1/\xi},\quad x>u.
    \end{eqnarray}
Wit suitable threshold $u$, we get $n_u$ excesses ($x_i -u$)'s leading to the maximum likelihood estimate of $\xi$ and $\widetilde\sigma=\widetilde\sigma(u) = \sigma + \xi(u-\mu)$. Here $(\mu,\sigma,\xi)$ is the triple of location, scale and shape parameter involved in the GEV model fitting the block maxima $M_n$ of an independent and identically distributed sample $X_1, \ldots, X_n\sim X\sim F$, i.e., 
\BQN\label{GEV}
\pk{M_n \le x} \approx G(x; \mu,\sigma, \xi):= \expon{-\left(1+\xi\frac{x-\mu}{\sigma}\right)_+^{-1/\xi}}
\EQN
holds under regular conditions. 

The exceedance probability  in \eqref{Fx} provides information on the amount of time a risk is expected to exceed certain performance levels, which is commonly used to predict extreme events such as floods, hurricanes and earthquakes.  In addition, one may invert \eqref{Fx} to obtain a high quantile of the underlying distribution or $T$-year return level stated below. 

For $T >1/\overline F(u)$, we have the $T$-year return level (i.e., the level $x_T$ that is exceeded on average every $T$ observations (with probability $p=1/T$) is  
\begin{equation*}
    x_T = u+\frac{{\widetilde\sigma}}{{\xi}}\left(\left(\frac{1}{T\overline F(u)}\right)^{-{\xi}}-1\right).
\end{equation*}
Thus, the minimum level that the potential risk is exceeded with small probability $1-q$, the so-called Value-at-Risk (VaR) at level $q$,  denoted by $\VaR_q(X)$,  is given by 
\begin{equation}\label{eq: VaR}
    \VaR_{q}(X) = u+\frac{{\widetilde\sigma}}{{\xi}}\left(\left(\frac{1-q}{\overline F(u)}\right)^{-{\xi}}-1\right),\quad q> F(u).
\end{equation}

\COM{According to the calculated $\VaR_q$ values, the spillover rate $E$ under each significance level $1-q$, which is used to test the degree of risk coverage, can be obtained by $E = n/N$, where $N$ is the sample size, $n$ is the frequency of excess over $\VaR_q$. If $E$ is larger than $1-q$, then the $\VaR$ model underestimates the risks, otherwise it is too conservative. Thus, the better model is indicated with smaller difference between $E$ and $1-q$. }

%\subsubsection{Conditional Value-at-Risk (CVaR)}
\CCL{$\VaR$ is commonly used as a risk measure to determine the potential capital premium due to its simplicity and back-testing properties. However, it does not reflect the tail magnitude but its probability. Alternatively, } conditional Value-at-Risk (CVaR) is the expected value of the losses exceeding the corresponding VaR, so it is represented as
\begin{equation*}
    \CVaR_p(X) = \E{X|X>\VaR_p(X)} = \VaR_{p}(X) + \E{X-\VaR_{p}(X)|X>\VaR_p(X)}.
\end{equation*}

\CCL{Since the threshold-excesses follow asymptotically generalized Pareto distribution, one may approximate the tail expectations as} %for any $u>u_{0}$, the mean excess function can be defined as:
\begin{equation}\label{eq: mep}
    \E{X-u|X>u} \approx \frac{\widetilde\sigma(u)}{1-\xi} =  \frac{\sigma + \xi(u-\mu)}{1-\xi}, \quad \xi<1.
\end{equation}
\COM{
Hence, it can be obtained that:
\begin{equation*}
    \E{X-\VaR_{p}|X>\VaR} = \frac{\sigma(u)+\xi(\VaR_{p}-u)}{1-\xi}.
\end{equation*}
When $\xi<1$,
}
Thus, 
\begin{equation}\label{eq: CVaR}
       {\CVaR_{p}}(X)={\VaR_{p}(X)} +\frac{\widetilde\sigma(\VaR_p(X))}{1-\xi} = \frac{{\VaR_{p}(X)}}{1-{\xi}}+\frac{{\sigma} - \xi \CCL{\mu}}{1-{\xi}}.
\end{equation}

One can obtain estimates of exceedance probability, $\VaR$ and $\CVaR$ using Eq. \eqref{Fx}, \eqref{eq: VaR} and \eqref{eq: CVaR} with  related parameters estimated and empirical survival probability $\widehat{\overline F}(u)={n_u}/n$ provided that $n_u$ exceedances over $u$ come from $n$ samples. 

\CCL{Besides the GEV and GP models the block maxima and threshold excesses aforementioned, a particularly elegant formulation of characterizing the extreme value behavior is derived from the theory of point processes (PP) which consider exceedances of threshold as events in time and model thus both the occurrence}  and intensity of exceedances. % can both be inferred in the PP model. 
Given a sequence of r.v. $X_i$'s with values in a state space $\mathcal A$, we define, for any set $A\subset \mathcal A$, the r.v. $N(A)$  is the number of points $X_i$ in the set $A$, which can formulize a point process under certain conditions. 
The intensity measure of this process \CCL{is one of the key summary features}, defined as
\begin{equation*}
	\Lambda (A) = \E{N(A)}, 
\end{equation*} 
which gives the expected number points in any subset $A$. The intensity density function is then defined by its derivative function if this exists with $A = \prod ^{d}_{i=1}\left [ a_{i},x_{i} \right ]$, i.e., 
\begin{equation*}
\lambda \left ( x \right ) = \frac{\partial \Lambda (A)}{\partial x_{1}\cdots\partial x_{d}}. 
\end{equation*}
\COM{For homogeneous Poisson point process, the parameter $\lambda$ is a positive constant, such that for $A = \left [t,s  \right ]\in T$, 

\begin{equation*}
    N(A)\sim Poisson (\lambda (s-t)).
\end{equation*}
}
\CCL{Let $X_1, \ldots, X_n$ be an independent and identically distributed sequence, which may consists of the observation of a potential risk $X$ satisfying \eqref{Fx} or \eqref{GEV}. Then the point process of $(i/(n+1), X_i)$ on $A = \left [ t_{1},t_{2} \right ] \times (u,\infty )\subset [0,1]\times \R$, denoted by $N(A)$, is given as
%a given threshold $u$ in time span $[t_1, t_2]\subset[0,1]$, i.e., 
$$N(A) := \#\{i\in\N, \ t_1\le i /(n+1)\le t_2,  \ X_i > u\},\quad A = \left [ t_{1},t_{2} \right ] \times (u,\infty ).$$
Under some weak conditions, we have $ N(A)\sim Poisson(\Lambda(A))$ holds asymptotically}
with intensity measure given as below. \begin{equation*}
    \Lambda(A) = \int _{(t,x)\in A} \lambda(t,x)\d t\!\d x
=(t_{2}-t_{1})\cdot \left [ 1+\xi \frac{u-\mu }{\sigma } \right ]^{-\frac{1}{\xi }}\cdot \I{1+\xi \cdot (u-\mu)/\sigma > 0},
\end{equation*}
where $\mu,\xi\inr$, and $\sigma >0$, which are the so-called location, shape, and scale parameter, respectively. $\I{\cdot}$ is the indicator function, \CCL{see Theorem 7.1.1 \citep{Coles2001}.} 
Maximum likelihood estimation is adopted here to estimate parameters involved in the PP model. PP log-likelihood, $\ell(\cdot)$, for a high threshold $u$, is given as
\BQN\label{eq: loglik}
    \lefteqn{ \ell\left ( \mu ,\sigma ,\xi ;x_{1},\ldots,x_{n} \right ) 
    }    \\
    &&= -n_{u} \ln \sigma  - \left ( \frac{1}{\xi} +1 \right )\sum_{i=1}^{n} \CCL{\ln}\left( 1 + \frac{\xi }{\sigma }\left ( x_{i}-\CCL{\mu}\right ) \right)\I{x_{i}> u}  
    - \left ( 1+\frac{\xi }{\sigma } \left ( u-\mu  \right )\right )^{-\frac{1}{\xi }}. \notag
\EQN
%where $n_{u}$ is the number of samples exceeding threshold $u$. 
Though this log likelihood function considers excesses, the parametrization is based on the GEV distribution function, \CCL{and thus is invariant to threshold. Consequently, the PP model can be adapted to allow for non-stationary effects to include temporal or covariate effects in the parameters with even time-varying thresholds} (\citep{Gilleland2016}).

\subsection{Multiple-Criterion Decision-Making Method (MCDM)} \label{app: MCDM}
Multiple-Criterion Decision-Making Method (MCDM) provides support for decision-makers to make the optimal choice among various alternatives and criteria. Here we consider two most useful methods of MCDM, namely, GRA and TOPSIS: 
\begin{enumerate}
    \item Grey relational analys (GRA) is commonly used to  tackle the problem of sample size and data uncertainty and achieve ranking objects by the relation degree to the reference object based on its {performance matrix} and {distinguishing coefficient}. 
    \item The  {Technique for Order Preference by Similarity to an Ideal Solution (TOPSIS)} can reduce the influence from the subject choice of distinguishing coefficient involved in GRA. In the application of TOPSIS, we consider  {Shannon entropy weights} to determine the decision matrix and model the relative distance of the Positive-ideal solution to the Negative-ideal solution. 
\end{enumerate} 
Both methods require the data normalization due to the different measurement of units and magnitude. 
In Section \ref{app: GRA} and \ref{app: TOPSIS}, we will illustrate how to give grey relational grade and closeness coefficient which will be used to rank the alternatives, see \eqref{GRG} and \eqref{CC} below.

\subsubsection{Grey relational analysis (GRA)}\label{app: GRA}
Once we have the performance information of $n$ alternatives on $p$ criteria, say $(x_{ij})_{n\times p}$, we shall select and combine the most representative data of each indicator as the reference sequence $x_{0}$. Grey relational coefficients are calculated between the reference sequence and all other comparability sequences. According to the calculated grey relational coefficients, grey relational grade can be achieved. Finally, if we sort the grade into a table, in this paper, the result with the highest grey relational coefficient will be the province most affected by the flooding damages. The specific description of steps are as follows.

\textbf{Step 1: Dimensionless processing (Normalization).}
Suppose $X_{j} = (x_{1j},x_{2j},\ldots,x_{nj})^\top,\ j=1,\ldots, p$ is the performance of $n$ alternatives on the $j$th criterion. To unify the indicators, normalization should be applied. We consider the normalization of “the larger the better” and "the smaller the better" criteria to standardize the data as follows.
\begin{equation} \label{normalizationGRA}
x_{i j}^{\prime}= 
\left\{
\begin{array}{ll}
\displaystyle\frac{x_{i j}-\min_{i} \left(x_{i j}\right)}{\max_{i} \left(x_{i j}\right)-\min_{i} \left(x_{i j}\right)}, & \mbox{for positive-correlated indicators,}\\
\displaystyle\frac{\max_{i} \left(x_{i j}\right)-x_{i j}}{\max_{i} \left(x_{i j}\right)-\min_{i} \left(x_{i j}\right)}, & \mbox{otherwise.}
\end{array}
\right. 
\end{equation}

Since the flooding risk vulnerability has positive correlation with disaster losses, that is, the higher the risk vulnerability, the greater the loss the province will face. Based on two parts in this paper, risk exposure indexes are positively correlated with the flooding risk vulnerability (the larger the risk exposure indicators, the higher risk vulnerability); however, the regional economic development level indexes show negatively correlation with the flooding risk vulnerability. Hence, the regional economic development level indexes are still needed to be inverted, normalizing the data according to  "the smaller the better". 
%to ‘the larger the better’.

\textbf{Step 2: calculate Grey relational coefficients.} 
\COM{After data preprocessing, the grey relational coefficients are calculated. Due to the previous inverted treatment of regional economic development indexes, all the indicators now display positive relationship with flooding risk vulnerability,}
Since the normalized data $x_{ij}$'s represent the scale of flooding risks, the largest data of each indicator can be taken as the most vulnerability representative, which are selected as a reference sequence $x_{0}: x_{0} = (x_{01},x_{02},\ldots,x_{0p})^\top = \vk 1$ in view of Eq. \eqref{normalizationGRA}. % According to our previous data normalization method, the reference sequence should be a sequence of 1. 
In the following, we calculate the grey relational coefficient to measure the \LE{similarity} of the $i$th alternative (province) performed on the $j$th indicator.
\begin{equation*}
\LE{\tau \left(x_{0j}, x_{i j}\right)} =\frac{\min _{i j}\left|x_{0j}-x_{i j}\right|+\xi \max _{i j} \left|x_{0j}-x_{i j}\right|}{\left|x_{0j}-x_{i j}\right|+\xi \max _{i j} \left|x_{0j}-x_{i j}\right|} = \frac{\LP{\zeta}}{\LE{\left|x_{0j}-x_{i j}\right|}+\LP{\zeta}}.
\end{equation*}
Here \LP{$\zeta$ represents distinguishing coefficient} ranging from 0 to 1; the smaller $\zeta$ indicates larger distinguishing power; 0.5 is taken by default as many other papers.

\textbf{Step 3: calculate  grey relational grade.}
Grey relational grade is the weighed sum of grey relational coefficients obtained in Step 2, which shows the level of similarity between reference sample $x_0$ and all other comparable objects $x_i$. Equal weight is considered by default. The formula of grey relational grade is as follows.
\begin{equation}\label{GRG}
\LE{\tau \left(x_{0}, x_{i}\right)}=\sum_{j=1}^{p} w_{j} \tau \left(x_{0j}, x_{ij}\right),\quad \mbox{where}\ \LP{w_1 = \cdots =w_{p}=1/p,\ p=13}.
\end{equation}
 The grade is in the range 0 to 1, the closer the grade to 1, the higher the vulnerability of the $i$th province.
 
 GRA will rank the alternatives according to the grey relational grade given in Eq. \eqref{GRG}. The larger the grade is, the higher vulnerability the alternative has. In the following, we consider an alternative approach to measure the vulnerability, namely, the TOPSIS.

 \subsubsection{{Technique for Order Preference by Similarity to an Ideal Solution (TOPSIS)}}\label{app: TOPSIS}
\COM{Before applying TOPSIS to give the rank of flooding risk vulnerability in 19 provinces, the weight for each indicator should be determined.}
In contrast with GRA, TOPSIS is free of distinguishing coefficient with a reasonable entropy distribution of weight for each criterion. Essentially, the ranking follows the relative distance to the extreme ideal solutions. This can be carried out following the steps below. 

\textbf{Step 1: calculate  entropy and entropy weight.}
Based on  the normalized data $x_{ij}'\in[0,1]$ using  Eq. (\ref{normalizationGRA}), we amplify it with a small extent $a = 0.01$ and proceed the coordinate translation as
    \BQN\label{normalizationTOPSIS}
    x_{ij}'' = x_{ij}'+a,\quad 
r_{i j}=\frac{x_{i j}''}{\sum_{i=1}^{n} x_{i j}''}, \quad i=1,2, \ldots, n;\ j=1,\ldots, p. %=13.
\EQN
Here, the amplitude $a$ added to the dimensionless data is to eliminate the effect of the logarithmic calculation. The selection of $a=0.05$ is motivated by  \citep{Sun2020} since then the resulted evaluation shows great significance. 

Now,  calculate the entropy for each criterion as
  \begin{equation*}
e_{j}=-\lambda \sum_{i=1}^{n} r_{i j} \ln r_{i j}\in[0,1], \quad j=1,2, \ldots, p,
\end{equation*}
where $\lambda=\frac{1}{\ln n}$, the Boltzman's constant. 
Finally, we obtain the \LE{entropy weight $w_j$} of the $j$th indicator as 
\begin{equation}\label{eq: entropy}
\LE{w_{j}} =\frac{\overline e_{j}}{\sum_{j=1}^{p}\overline e_{j}}, \quad j=1,2, \ldots, p,
\end{equation}
where $\overline e_{j}= 1-e_j$ models the degree of diversification $e_j$'s of the information. 

Based on the calculated weight for each indicator, TOPSIS is applied to help determine the ideal scheme, that is the province with the highest and lowest %least
flooding risk vulnerability in this paper.

{\bf Step 2: construct  weighted normalized decision matrix $V$.}
\begin{equation*}
R=\left(\begin{array}{ccc}
r_{11} & \cdots & r_{1 p} \\
\vdots & \ddots & \vdots \\
r_{n1} & \cdots & r_{np}
\end{array}\right), \quad \LE{r_{i j}} = \left\{
\begin{array}{ll}
  \displaystyle \frac{x_{i j}}{\max _{i}\left(x_{i j}\right)}  ,&\mbox{positive-correlated indicators},\\
    \displaystyle\frac{\min _{i}\left(x_{i j}\right)}{x_{i j}}  ,&\mbox{otherwise}.
\end{array}
\right.
\end{equation*}
Thus, we obtain the weighted {normalized} decision matrix $V$.
\begin{equation*}
V=\left(v_{ij}\right)_{np}, \quad \mbox{with}\ v_{i j}=\LE{w_{j}} r_{i j},
\end{equation*}
where the weights $w_j$'s are given by \LE{entropy method} in Eq. \eqref{eq: entropy}.

 {\bf Step 3: calculate the positive-ideal solution (PIS) and the negative-ideal solution (NIS). }
      \begin{equation*}
\left\{
\begin{array}{ll}
A^{+}=\left\{v_{1}^{+}, v_{2}^{+}, \ldots, v_{p}^{+}\right\},& v_{j}^{+}=\max _{i}\left(v_{i j}\right),
 \\
A^{-}=\left\{v_{1}^{-}, v_{2}^{-}, \ldots, v_{p}^{-}\right\}, & v_{j}^{-}=\min _{i}\left(v_{i j}\right).
\end{array}
\right.
\end{equation*}

{\bf Step 4: calculate the distance of each object from PIS and NIS.}
        \begin{equation*}
S_{i}^{+}=\sqrt{\sum_{j=1}^{p}\left(v_{i j}-v_{j}^{+}\right)^{2}},
\quad 
S_{i}^{-}=\sqrt{\sum_{j=1}^{p}\left(v_{i j}-v_{j}^{-}\right)^{2}}.
\end{equation*}
 {\bf Step 5: calculate the closeness coefficient of each object.} 
    \begin{equation}\label{CC}
C_{i}^{*}=\frac{S_{i}^{-}}{S_{i}^{+}+S_{i}^{-}}, \quad 0<C_{i}^{*}<1.
\end{equation}

\subsection{\textbf{A Pricing Scheme of Catastrophe Bond.}} 
Trigger model, payoff function, principal wipe out function and accrued coupon should be determined prior to pricing measurement since coupon payment and  \CCL{principal} redemption might be changed once catastrophe (CAT) event occurs.

Consider a trigger process $\{Y_{t}, \ t\ge0\}$ as a component-wise non-decreasing, non-negative, and right continuous stochastic process 
\BQN\label{eq: trigger}
Y_{t}= f(X_{1},X_{2},\ldots,X_{N_{t} } ), \quad t\ge 0, 
\EQN
where $ \left\{ X_{n},n\in \N \right\}$ is a sequence of  random variables \CCL{modelling precipitation severity} and  $\left\{N_{t},t\ge 0  \right \} $is a counting process to model the occurrence of floods, and $f$ is a component-wise non-decreasing function.
\\
There are various kinds of trigger $Y$, for instance 
\begin{itemize}
    \item[(i)] The aggregate amount of loss due to the flood;
\item[(ii)] The number of earthquakes which consider both magnitude and frequency as below
\begin{equation*} 
Y_{t}=\sum_{ j=1}^{N_{t}}  \I{X_{j}> u}, \quad u\ge 0,
\end{equation*}
where $u$ is a high threshold and $N(t)$ is the number of the exceedances over the threshold observed in $[0,t]$.
\end{itemize}

Next, the pay off function $\Pi(y): [0, \IF) \mapsto [0,1]$, is a non-increasing function linking the natural disaster risks via the trigger process $Y_t = f(X_1, \ldots, X_{N(t)})$ and the financial securities, 
stimulate a plan of allocating the  \CCL{principal} between the investor and sponsor according to the development of triggering process. Specially, the time of  \CCL{principal} wiped out is denoted by
\begin{equation*} 
\tau  = \inf \left \{ t\ge0:\  \Pi (Y_{t} )=0 \right \}.
\end{equation*}
Next, we introduce the pricing measure as a product of the natural disaster risk and financial risk. The pricing at time $t$ is essentially the discounted value of future coupon payments plus the discounted value of the remaining  \CCL{principal}.
\begin{eqnarray}\label{price}
\qquad P_{t}&=&K\EE_{t}^{Q_{1}\times Q_{2}}  \left[ \sum_{s=\left \lfloor t \right \rfloor+1 }^{\left \lfloor \tau   \right \rfloor \wedge T} D(t,s)(R+i_{s})\Pi (Y_{s-1})+
D(t,\tau )\vartheta 1(\tau \le T) + D(t,T)\Pi (Y_{s-1})\right]\\
&=& K\EE_{t}^{Q_{1}}\left\{  \sum_{s=\left \lfloor t \right \rfloor+1 }^{\left \lfloor \tau   \right \rfloor \wedge T} \Pi (Y_{s-1})\EE_{t}^{ Q_{2}}[D(t,s)(R+i_{s})] \right \}  + K\EE_{t}^{Q_{1}}[\Pi (Y_{T})]\EE_{t}^{ Q_{2}}[D(t,T)], \nonumber \\ \nonumber
&& + K\EE_{t}^{Q_{1}}\left \{(\tau  - \left \lfloor \tau   \right \rfloor) \Pi (Y_{\left \lfloor \tau  \right \rfloor  })\I{(\tau \le T)} \EE_{t}^{ Q_{2}}[D(t,s)(R+i_{\tau  })|\tau ]  \right \},  \nonumber
\end{eqnarray}

where the pricing measure $Q_{1}$ is the distorted probability measure of the catastrophe disaster risk $X$, $\EE_{t}^{Q_{1}} [\cdot ]$ terms usually require simulation to evaluate. 

The performance of financial market is reflected by the interest rate process ${r_{t}, t\ge 0}$ and the Shibor process ${\ell_{t},t\in N}$ using a risk-neutral pricing measure $Q_{2}$ for the arbitrage-free financial market according to the well-established APT to price the interest rate risk. We link the financial market risks with Eq. \eqref{price} through
$$D(t,T) = \expon{-\int_t^T r_t dt},\quad i_t = e^{\ell_t}-1.$$

We model $\{(r_t,\ell _{t}),t\ge 0\}$ as a bivariate correlated Ornstein-Uhlenbeck (OU) process (i.e., Vasicek models \citep{Rogemar2004}) under the risk- neutral measure $Q_{2},$ denoted by
\begin{eqnarray}\label{eq: financial risk}
\left\{
\begin{array}{l}
 \d r_{t}=a_{r}(b_{r}-r_{t})\d t+\sigma _{r} \d W_{r,t},   \\   
  \d \ell_{t}=a_{\ell}\left(b_{\ell}-\ell_{t}\right) \d t+\sigma_{\ell} \d W_{\ell, t}, 
\end{array}
\right.
\end{eqnarray}
where $a_{\cdot}, b_{\cdot}$ and $\sigma _{\cdot}$ are positive numbers corresponding to the rate of mean reversion, the long run mean, and  the volatility, accordingly. And $W_{r,t}$ is a standard Brownian Motion under $Q_{2}$.. Here the two Brownian motion processes are constant correlated, i.e., for some for some $\rho \in \left (- 1, 1 \right )$,
$$ dW_{\ell, t}dW_{r, t}=\rho d t, \quad t \geq 0.$$ 

\cL{It follows from the appendix of \citep{tang_yuan_2019}, we have (see also \citep{Rogemar2004})}
\begin{equation}
\label{etqdts}
\EE_{t}^{Q^{2}}[D(t, s)]=A(t, s) e^{-B(t, s) r_{t}}, \quad s \geq t \geq 0,
\end{equation}
\cL{where}
\begin{equation*}
\left\{
\begin{array}{l}
 \displaystyle A(t, s)=\exp \left\{\frac{(B(t, s)-(s-t))\left(a_{r}^{2} b_{r}-\sigma_{r}^{2} / 2\right)}{a_{r}^{2}}-\frac{\sigma_{r}^{2} B(t, s)^{2}}{4 a_{r}}\right\}, \\
 \displaystyle B(t, s)=\frac{1-e^{-a_{r}(s-t)}}{a_{r}}.
\end{array}
\right.
\end{equation*}
And 
$\EE_{t}^{Q^{2}}\left[D(t, s) i_{s}\right] $ can be calculated by
\begin{eqnarray}
\label{etqdtsi}
\EE_{t}^{Q^{2}}\left[D(t, s) i_{s}\right] &=& \EE_{t}^{Q^{2}}\left[D(t, s) e^{\ell_{s}}\right]-\EE_{t}^{Q^{2}}[D(t, s)] \\
&=& \tilde{A}(t, s) \exp \left\{-B(t, s) r_{t}+\tilde{B}(t, s) l_{t}\right\}-A(t, s) e^{-B(t, s) r_{t}}, \quad s \geq t \geq 0,\notag
\end{eqnarray}
where $A(\cdot,\cdot)$ and $B (\cdot,\cdot)$ are defined as above, and  $\tilde{A}(\cdot,\cdot)$ and  $\tilde{B}(\cdot,\cdot)$ are introduced as below.
\begin{equation*}
\left\{
\begin{array}{l}
\tilde{A}(t, s)=\exp \left\{-\left(C_{1}(t, s)+C_{2}(t, s)\right)\right\},\\
\tilde{B}(t, s)=e^{-a_{\ell}(s-t)}
\end{array}
\right.
\end{equation*}
with

$$
\left\{
\begin{aligned}
C_{1}(t, s)=&\left(b_{r}-\frac{\sigma_{r}^{2}}{2 a_{r}}\right)(s-t)+\frac{3 \sigma_{r}^{2}}{4 a_{r}^{2}}+\frac{\rho \sigma_{r} \sigma_{\ell}}{a_{\ell}\left(a_{\ell}-a_{r}\right)}+\frac{\sigma_{\ell}^{2}}{4 a_{\ell}}+b_{\ell}-\frac{b_{r}}{a_{r}}, \\
C_{2}(t, s)=& \frac{\sigma_{r}^{2}}{4 a_{r}^{2}} e^{-2 a_{r}(s-t)}+\left(\frac{b_{r}}{a_{r}}-\frac{\sigma_{r}^{2}}{a_{r}^{2}}\right) e^{-a_{r}(s-t)}+\left(\frac{\rho \sigma_{r} \sigma_{\ell}}{a_{r} a_{\ell}}-b_{\ell}\right) e^{a_{\ell}(s-t)} \\
&-\frac{\rho \sigma_{r} \sigma_{\ell}}{a_{r}\left(a_{\ell}-a_{r}\right)} e^{\left(a_{\ell}-a_{r}\right)(s-t)}-\frac{\sigma_{\ell}^{2}}{4 a_{\ell}} e^{2 a_{\ell}(s-t)}.
\end{aligned}
\right.
$$

\bibliographystyle{plain}
\bibliography{ref.bib}
\end{document}